\documentclass[letterpaper]{article} 
\usepackage{aaai2026}  
\usepackage{times}  
\usepackage{helvet}  
\usepackage{courier}  
\usepackage[hyphens]{url}  
\usepackage{graphicx} 
\usepackage{natbib}  
\usepackage{caption} 
\usepackage{algorithm}
\usepackage{algorithmic}
\usepackage{xcolor}
\definecolor{traditional-color}{HTML}{f3d0ee}
\definecolor{operationalist-color}{HTML}{c7eefb}
\definecolor{free-agent-color}{HTML}{c3f1c8}
\newcommand{\tradcolor}[1]{\colorbox{traditional-color}{#1}}
\newcommand{\opcolor}[1]{\colorbox{operationalist-color}{#1}}
\newcommand{\freecolor}[1]{\colorbox{free-agent-color}{#1}}

\usepackage{newfloat}
\usepackage{listings}
\DeclareCaptionStyle{ruled}{labelfont=normalfont,labelsep=colon,strut=off} 
\floatstyle{ruled}
\newfloat{listing}{tb}{lst}{}
\floatname{listing}{Listing}
\title{Locating Translation as a Craft in the Age of AI}
\author{
    Daniel Chechelnitsky\textsuperscript{\rm1},
    Sireesh Gururaja\textsuperscript{\rm1}\equalcontrib,
    Seyi Olojo\textsuperscript{\rm2}\equalcontrib,
    Giuseppe Attanasio\textsuperscript{\rm3},
    Wesley Hanwen Deng\textsuperscript{\rm4},
    Chrysoula Zerva\textsuperscript{\rm3, 5},
    Maarten Sap\textsuperscript{\rm1}
}

\affiliations{
    \textsuperscript{\rm 1}Carnegie Mellon University, Pittsburgh, PA, USA\\

    \textsuperscript{\rm 2}University of California, Berkeley, Berkeley, CA, USA\\
    \textsuperscript{\rm 3}Instituto de Telecomunicações, Lisbon, Portugal\\
     \textsuperscript{\rm 4}Microsoft Research, New York, NY, USA\\
     \textsuperscript{\rm 5}ECE - National Technical University of Athens, Athens, Greece\\
    dchechel@andrew.cmu.edu
}

\usepackage{bibentry}

\begin{document}

\maketitle

\begin{abstract}
Rapid development of Large Language Models (LLMs) and similar automated approaches for translation tasks is increasingly affecting the landscape of translation technologies. 
As concerns about the outsourcing of translator work to these automated translation tools grow, it is increasingly crucial to gather insights from the translation community directly.
To this end, we conduct an interview study with 19 professional 
translators working across 11 languages and 11 domains to understand their perspectives, experiences, and concerns with using translation technologies in their work.
We find that 
translators are cautious when incorporating new tools into their workflow, with several expressing concerns that machine translation (MT) and LLMs are infringing on the necessary human aspects and verification processes of translation. Importantly, translators are worried that these tools have potential for harmful downstream effects due to compromising the human aspects of translation work.
These findings demonstrate the need to develop translation technologies that directly serve translators' needs rather than replacing human translation. This can be done by focusing more on the assistive tools that emphasize the uncertain, social, and ultimately human character of translation, rather than automation.
\end{abstract}


\section{Introduction}

\textit{Translation technologies} are tools that assist users with translation tasks \cite{alcina2008translation}. However, as technologies like Computer Assisted Translation (CAT) tools and Large Language Models (LLMs) have become more widely used in professional translation tasks \cite{youdale2022computer, zaki2024revolutionising}, the question emerges of what effects these tools have on translators' work. Given the ongoing rapid development and user accessibility, narratives of technological progress and general capability \cite{eloundou_gpts_2024}, and the manner in which these tools present translation as a straightforward mapping of input words to output words, translation is currently at risk of premature attempts at automation. This exposes professional translators to uncertainty and economic precarity, similar to professional artists and musicians \cite{jiang2026professional, chu2022empirical}. There is great concern for what will happen if this trend of transition fully replacing the human continues \cite{merchant2025ai}.




Prior work in translation technologies has shown limitations of state-of-the-art translation tools. For instance, Computer Assisted Translation (CAT) tools, which are specialized word and phrase tools with specialized glossaries called \textit{translation memories}, are still limited to hyper-specialized translation tasks \cite{alotaibi2014teaching, vukalovic2021analysis}. This is due to how translation tasks vary with regard to source-target language pairing, domain, and intended audience. Contrastingly, while the increased usage of LLMs provides a more generalizable translation ability, they often also show decreased performance and increased risk of hallucinations \cite{mohammed-niculae-2025-context, zhang-etal-2025-good, tang-etal-2025-mitigating}.

In all this, an overlooked aspect of developing useful translation technologies responsibly is the translator expertise. While prior work has outlined the need for translator perspectives in the development pipeline \cite{liebling-etal-2022-opportunities, carpuat-etal-2025-interdisciplinary}, there is still limited work in engaging with the expertise of professional translators. This is because translation goes beyond simple mapping of words in one language to the other: it is an inherently context-dependent task that requires linguistic and domain expertise \cite{diaz_what_2024}.



In this work, we investigate professional translators' experiences with and attitudes about translation technologies and their affordances, a first-of-its-kind study, looking into the following research questions:
\begin{enumerate}
\item[\textbf{RQ1}] \textbf{What are the goals of translation and what expertise is required to achieve said goals?}

\item[\textbf{RQ2}] \textbf{What are the affordances of current translation technologies?}

\item[\textbf{RQ3}] \textbf{What are translator attitudes and concerns towards translation tools?}
\end{enumerate}



To address these questions, we performed semi-structured interviews with 19 trained translation professionals across 11 domains and 11 languages.
Through thematic analysis, our results show that translators in our study tend to encourage using both CAT and AI tools, while simultaneously arguing that the tools should only be used by a professionally trained human for translating the specific task and domain (Section \ref{sec:human-discussion}). 
This ensures that translators have access and can validate any errors the tools may have made. 
However, there is also a growing distrust of LLMs and AI tools among some professional translators due to ethical and economic concerns such as AI infringing on individuals' expertise (Section \ref{sec:infringing-discussion}).
Lastly, there is a growing concern that as the translation technology landscape keeps changing, the role of a translator will be minimized, with the added risk of the creative and human role of a translator being automated away. 
It is therefore crucial to emphasize the required role of human translators within translation technology. 
Building on these findings, we discuss the nuances and implications of using these tools for translation and what it means for translators' career longevity (Section \ref{sec:craft-erasure}).
\section{Background}


In this section, we scope the problem within the context of translation technology, largely based on the human-centered machine translation tools described in the work of ~\citet{liebling-etal-2022-opportunities}. To do this, we describe terminology, provide concrete examples, and discuss current benefits and oversights for translation technologies and tools.

\subsection{The Current Landscape of Translation Tools}


We broadly define translation technologies as tools that assist users with translation tasks. These users might be lay or professional, and the types of tasks they perform with these technologies can vary significantly. Generally, however, we focus on three types of translation technology in this work, in ascending order of specificity. The first are general-purpose tools aimed at a broad audience, in which translation is one function of many. This category is exemplified by LLM-based systems such as ChatGPT. Next are machine translation tools, which map input sentences to output sentences, like Google Translate or DeepL. Finally, tools that are built for professional translators to assist in translation without performing it, such as computer-aided translation (CAT) tools. These tools are often designed to center the needs of large-scale translation projects, prioritizing translation consistency. 




\subsection{History of Translation Technologies}


The goal of translation aided or performed by computers has a long history. The archives of the European Association for Machine Translation, for instance, list correspondence from as early as 1947 discussing the possibility of translating text by computer to aid with the mission of UNESCO \citep{weaver_weiner_letter}. Machine translation was also the primary subject of the infamous 1966 ALPAC report \citep{alpac1966}, which found that the quality of both machine-only and post-edited translation was insufficient, and thereby, in many accounts, led to the first AI winter, in which funding for MT work in the U.S. was severely curtailed \citep{hutchins_alpac_2003}. The report itself, however, notes that while full automatic translation was infeasible at the time, computer-aided translation merited further development, with two case studies of computer-aided translation in Europe whose function mirrors the translation memories of CAT tools today. MT has since recovered and retained a strong research community, leading to the development of foundational AI techniques like attention \citep{bahdanau2014neural} and the transformer architecture \citep{vaswani2017attention}.

\section{Related Work}
Here, we discuss ongoing parallel work in human-AI interaction, specifically emphasizing expert experiences. Our work within the translation technology space similarly provides novel insights into the field.


\subsection{Characterizing Expert Work Processes}

Prior to attempting to understand the role of AI in reshaping work practice, we argue that it is important to understand the work practice at risk of being reshaped. We therefore draw on literature that characterizes experts' conceptualizations of their own work through semi-structured interviews.~\citet{10.1145/3777475}, for instance, describe the work processes of professionals who produce audio descriptions of video content, and \citet{gururaja-etal-2025-beyond} characterize the processes of knowledge workers engaged in research in document corpora. We extend this approach to translators to understand how they conceptualize their own work, and whether the technologies that aim to either augment or replace them operate on compatible representations of that work. 

Recent work from Human-Computer Interaction has also examined how machine translation is being used in high-stakes scenarios such as medical domains \cite{deng2022beyond, mehandru2023physician, robertson2021three}, as well as providing guidelines for designing more
human-centered machine translation systems \cite{liebling-etal-2022-opportunities, robertson2021three, carpuat-etal-2025-interdisciplinary, savoldi-etal-2025-translation}. Studies like these further our understanding of the fields that new technologies seek to disrupt. Our work expands on this by exploring whether these technologies can in fact address the concrete needs of expert users. As AI becomes ubiquitously available, it is important to understand how expert workers perceive the technology, and how it may or may not assist with their work. \citet{Woodruff2023HowKW}, for instance, find that while knowledge workers see some utility in AI tools, they largely see them as tools to "perform menial work, under human review", and view them as exacerbating existing trends like deskilling in their industries.
Similarly, \citet{yu_why_2025} focus on when and why decision makers in organizations might choose to use AI or not, finding that AI tool use is often impacted by factors outside of its task performance, including the legal or reputational consequences for its use.

\subsection{Expert Perceptions of AI Tools} 

More specifically to translation, projects like Translators Against the Machine \citep{minshall_translators_2025} and Blood in the Machine \citep{merchant2025ai} have collected stories from translation professionals whose work has been impacted by translation technologies, showing that while translators' opinions of AI as a technology vary, their perception of AI as an instrument of social change focuses on its use to de-value their work, with some translators finding it economically infeasible to continue in the profession. 


Individual translators' perceptions of AI as a harbinger of precarity also reflect concerns about the broader impacts of the technology. A U.K. survey of 787 members of a trade organization for translators, for instance, found that 36\% of them reported job losses from the use of AI~\citep{soa_survey_2026}. In their work, ~\citet{seyi} argue that the naive use of primarily English-centric translation technologies reproduces colonial structures of harm. 

We therefore orient our work towards understanding how translators approach their work, what they consider valuable in it, how machines may or may not be in conversation with those values, and the broader implications for both translators themselves and the users of translated language.

\section{Methods}
We conducted semi-structured interviews to help provide more nuance and acute understandings of technological ways that these translation technologies should be contextualized \cite{gururaja-etal-2025-beyond, seyi}. 
We first conduct our interviews, then perform three independent rounds of open coding to conduct thematic analysis to extract commonalities and broader themes from our interviews \cite{braun2006using}.

\begin{table}[h]
\centering
\begin{tabular}{l c r}
\hline
ParticipantID & Age & Translation Domain \\
\hline
1        & 35-44       & Advertisement, News   \\
2        & 65+       & Academic     \\
3        & 25-34       & Entertainment, News     \\
4        & 25-34       & Freelance     \\
5        &   25-34     &  Academic, Freelance    \\
6        & 65+       & Academic, Literary     \\
7        &   25-34     &  Freelance   \\
8        &   35-44    &   Academic, Legal   \\
9        &    25-34    &   Freelance   \\
10        &   35-44     &  Academic   \\
11        &   25-34    &   Freelance   \\
12        &    25-34   &   Freelance   \\
13        &   65+     & Legal     \\
14        & 25-34       & UX-Design     \\
15        & 25-34       & Advertisement    \\
16        & 25-34       & Literary, Religious     \\
17        & 65+       &  Academic, Literary   \\
18        & 25-34       & Medical     \\
19        & 25-34       & Academic, Literary     \\
\hline
\end{tabular}
\caption{Interview Participants Translation Age and Domain}
\label{tab:age and domains}
\end{table}

\begin{table}[h]
\centering
\begin{tabular}{l c r}
\hline
Language & Count  \\
\hline
Spanish        & 7     \\
Swahili      & 3       \\
German        & 2       \\
French        & 2       \\
Russian        & 2        \\
Mandarin Chinese      & 2        \\
Arabic       & 2        \\
Korean       & 1        \\
Norwegian & 1 \\
Sanskrit        & 1         \\
Bengali       & 1         \\

\hline
\end{tabular}
\caption{Distribution of Interview Participants' Languages}
\label{tab:languages}
\end{table}

\paragraph{Recruitment}
To represent as many possible translators' viewpoints as possible, recruiting was done more broadly, regardless of specific translation language pair or translation domain. With an IRB obtained from the institution of the first author, recruitment of interviewees was conducted using convenience and snowball sampling stemming from the first author's academic and personal network \cite{etikan2016comparison, goodman1961snowball}. The first author additionally recruited through social media posts on LinkedIn, X, Bluesky, Facebook, and Reddit. All participants had self-selected themselves as translation experts prior to completing the consent script, and participants were recruited until experimental saturation was met \cite{guest2006many}. At the end of the recruiting process, 19 interviews went forward. Each interview lasted approximately 45 minutes, and upon completion, the interviews were de-identified for any participant and interviewer personal information.

 
\paragraph{Participants}

The translators covered a range of 11 different languages and 11 different domains (see Table \ref{tab:age and domains} and Table \ref{tab:languages}). These interviewees represented translators who have been translating for over 60 years, to those who were relatively new and have only been translating after LLMs became widespread in 2022. These languages include high-resourced languages like Spanish, Arabic, and Chinese, as well as lower-resourced languages like Swahili and Bengali. In total, 19 interviews were conducted by three of the authors between January and February 2026 over Zoom.

\paragraph{Qualitative Analysis Procedure}
For our results, we wanted to extract commonalities from what the interviewees expressed, while also providing nuance across the translators' specific domains and languages.  
To evaluate this, we used an inductive interpretivist approach \cite{thomas2003general} to synthesize our findings within existing literature for our research questions, as well as a grounded theory-informed approach where we extracted themes from our participant responses \cite{charmaz2006constructing}. In doing so, we obtained rich descriptive information on how these tools are usedm similar to related qualitative work \cite{seyi, gururaja-etal-2025-beyond}. The themes were drawn from codes that were obtained using a thematic analysis approach consisting of coding of in-vivo examples from the translation tasks described by the interviewees \cite{braun2006using}. Specifically, three authors completed these three rounds of coding:

\begin{enumerate}
    \item \textbf{Open Coding}: Each interview is coded individually by separate authors, with each author keeping an individual codebook.
    \item \textbf{Closed Coding}: The three authors who were closed coding met and discussed the common codes that emerged, and began merging existing codes. The final list included a codebook of 63 codes.
    \item \textbf{Axial Coding}: Two of the three authors from the first two rounds and an additional author performed axial coding by sorting the codes into sub-themes and themes. This follows the approach in thematic analysis \cite{braun2006using}. At the end, each research question was directly addressed using two to three themes.
\end{enumerate}

In addition to these codes, the first-author wrote down major takeaways of each participant as personal memos. These, in tandem with the codes and themes, provided organization and structure for discussing the results. Our findings therefore incorporate both broader motifs, as well as rich context-specific details, all based on the experiences of our interviewees.

 \section{Findings}

Based on our themes extracted from the qualitative open-coding \cite{braun2006using}, we found translators have strong personal opinions about state-of-the-art translation technologies.

We first describe our participants, their domains, and three emergent translator profiles. Then, following our research questions we first describe the translators themselves, the translation process and the human component of translation \textbf{(RQ1)}. We then discuss the tools they use, and how they fit into the current landscape of translation technologies \textbf{(RQ2)}. Finally, we discuss how translators incorporate the tools within their workflow and the general translator attitudes toward these tools \textbf{(RQ3)}. This includes how the affordances of translation tools impact the ways translators conduct their work, including what current limitations and restrictions they face. 



\subsection{Emergent Translator Profiles}

The 19 translators interviewed spanned domain expertise in academic literary translation, to document translation for a translation agency in medical, legal, and design domains, to freelance work (Table \ref{tab:age and domains}). 

Our translator pool is also heterogeneous in skills, prior experience, and preferences. All translators, regardless of age, have concrete viewpoints on what translation `should' be, and how it should be conducted. Whereas some translators might be more hesitant to use ML and AI tools, other translators who still have more flexible workflows are able to experiment and potentially incorporate newer translation technologies in their work.

Based on the translator roles and tool usage, we identify three main translator profiles to more helpfully characterize these translator experiences and approaches:
\begin{itemize}
    \item \tradcolor{\textbf{Traditionals (P2, P6, P8, P10, P13, P17, P19):}} Translators who became translators through academic or more traditional training. Preferred tools include dictionaries, pen and paper, and prior knowledge of a specific domain. Their work is also done more directly with clients. 
    \item \opcolor{\textbf{Operationalists (P3, P4, P5, P7, P11, P14, P15, P18):}} Individuals who still have a more traditional training for translation, often involving extensive schooling, but have used a wide variety of tools in their training, including CAT tools, MT, and even more recently LLMs. Their work is typically in collaboration with other translators or as part of an agency or corporation. Similarly, their work is often bounded, quite strictly, to CAT or other tools specifically prepared for them.
    \item \freecolor{\textbf{Free Agents (P1, P9, P12, P16):}} These translators have come to translation from a wide variety of domains, but are not working under a corporation or doing the job for anyone else except themselves. Typically have some sort of formal training via a degree or certificate, but are also willing to pick up new skills along the way. Often times they are online translators, or literary translators who work independently. The tools they use depend on individual preference.
\end{itemize}

Using these categories, we help communicate why certain translators tend to have certain preferences and approaches when it comes to using translation tools, going beyond just describing the translator's individual language pair and translation domain.

\subsection{Translation as a Craft}
\label{sec:decisions}

In this section, we address \textbf{RQ1} and describe the experiences of translators and the emergent themes that arise from the human component that is required in translation. These themes often are rooted in translators' beliefs and opinions formed from practicing their translation.

\subsubsection{Translators want to do a good job.} 
\label{sec:good-job}
Through our interviews, our participants communicated to us repeatedly the degree of care they exercised in translation in the pursuit of translation quality. Translators across domains and regardless of demographics often spoke about prioritizing the experience of the reader, in order to convey to them particular aspects of the text. Translators often balanced competing priorities in order to achieve what they saw as a high quality translation: for instance, sacrificing tone and register in legal documents (\tradcolor{P2}), or syntax from grammatically different languages to be legible (\tradcolor{P6}). Similar to how different translation domains reflect different needs and communicative intents, the definition of quality varied from translator to translator, though the focus on achieving it did not. 

\subsubsection{Translation requires preparation and cultural understanding.}
\label{sec:translation-requires-preparation}

A common practice among our participants across categories was a preparatory read-through of a text before beginning translation. \colorbox{free-agent-color}{P1},\tradcolor{P2},\opcolor{P3},\opcolor{P4},\opcolor{P5},\freecolor{P12}, and \tradcolor{P19} all described their process of familiarizing themselves with a text before beginning a translation, in order to understand the aspects of the text: technical terms or idiomatic constructions, the register and tone of the language, or even the "feel" of the language. 

\colorbox{traditional-color}{P19} states that both research and background knowledge are necessary to consider author style when translating. They describe their process for doing research as such:

\begin{quote}
\textit{`I will do some searching online to familiarize myself with the author, with their context. If there are any particularities about their own writing style that I should take into account while trying to translate the text.'}

\hfill ---P19, Traditional
\end{quote}

\colorbox{free-agent-color}{P1} also notes that on top of familiarity with the text, it is important to know how it fits into the surrounding culture, and describes language and culture as inherently linked. \colorbox{traditional-color}{P10} agrees and adds that the translator's role in translating is not to simply facilitate translation, but to include context to more accurately engage with the culture, place, and time as important components to have in consideration before proceeding with a translation.

Finally, \colorbox{operationalist-color}{P3} mentions that part of preparation for structured translation work requires knowing the audience. \colorbox{traditional-color}{P19} similarly echoes how for academic contexts this is critical because often times the language used is context specific, where one French word in one historic setting can mean a completely different thing in another.

\subsubsection{Translation has many possible goals.}
Though all of our participants identified as translators, the goals of this "translation" work were conceptualized differently. Legal and medical translators (\tradcolor{P13}, \opcolor{P4}, \opcolor{P18}), for instance, shared a deep concern for the consistency and literal accuracy of their work. Translators working in marketing copy and literary translation thought deeply about the effect their work had on the reader; \freecolor{P1}, for instance argued \textit{`if you're writing about meditation, it has to be meditative'}. Translators working on localization for software had the unique concern of the spatial characteristics of text: localized text still had to fit within the UI being translated. 

All of these goals were informed by a consideration beyond the content of the text to be translated: they required evaluating the context, audience, and intended effect of the translation in the reader, and each of those factors varied based on the kind of translation being performed. While participants like \tradcolor{P13} argued for precision in the legal setting, \tradcolor{P6}, a literary translator, argued that \textit{`style should not be sacrificed for
accuracy or idiomaticity'}. Notably, even within the same kind of translation, these goals may be contradictory, or as \tradcolor{P6} describes it, \textit{`being the servant of two masters'}. \tradcolor{P6} goes on to describe that while part of their goal is to translate the distinctive character of an author's voice (\textit{`when you translate fiction, Tolstoy or Dostoevsky, you don't want it to sound like Dickens or Thackeray.'}), that can be impossible to do that literally, as in the case of one author, whose syntax they described as \textit{`bulky, and totally unacceptable...really, to any English reader.'}.

Translation in these contexts therefore requires judgment on the part of the translator to make choices about how to prioritize factors like literality, readability, and mood in the service of the translation's goal. Rather than there being a "correct" translation, different translators may choose to prioritize different aspects of the text, in order to achieve different communicative effects. As \tradcolor{P19} puts it, \textit{`in the academic context, a good translation is one that is able to render the general sense and the argument of the original article, as well as the style, the original style of writing. But not by going word for word, and not at the expense of the understanding of the reader.'}

The choices required of translators can also extend to creative and artistic ones. \tradcolor{P6}, \tradcolor{P10}, and \tradcolor{P19} all describe cases where the source language can contain social context, like information about class or the social relations between characters, e.g. in the case of characters in Charles Dickens' novels that speak in the Cockney dialect, indicating a working-class background. In these cases, translators must make choices about how, or whether, to translate these cultural signifiers. \tradcolor{P19} discussed translating absurdist theater such as: \textit{`maybe a more fun thing to do would be to translate the play in an adaptation... moving the whole play to an American context so that the jokes land more accurately in English'}. \opcolor{P14} spoke of similar choices in brand localization that reflect different cultural priorities: \textit{`let's say if our brand is playful. Yeah. We use very direct wordplay in English, whereas in Korea, we might use… friendly ending particles or emojis'}.

\subsubsection{Translation expertise is acquired through time and experience.}


Our participants often emphasized the role of experience in their work. For example, while almost all of our participants mentioned using dictionaries, multiple participants emphasized that dictionaries still require judgment to use correctly. \tradcolor{P10}, for instance, recounts that \textit{`[students] will tell you what the first meaning [in the dictionary] is, and it'll be wrong because the dictionaries are just not very, well written...dictionaries, they don't provide meanings that are contextually appropriate for a particular genre or for a particular time period.'} \freecolor{P1} similarly discusses the nuance of using Norwegian-Norwegian dictionaries to establish the nuances of a word's meaning before going to a Norwegian-Russian dictionary: \textit{`First choice is...a Norwegian-Norwegian dictionary, to understand if it's a word that I'm not familiar with...I need to understand how it fits in.  Then, I look at the Norwegian-Russian dictionary.'}

Multiple participants also emphasized that developing skill in translation was both difficult and necessary, and argued against the use of tools like LLMs or even MT during that skill development period. \tradcolor{P13}, for instance, says that \textit{"I wouldn't recommend it for a beginner...you have to know what you are doing"}, arguing further that it threatens the development of a translator's \textit{Weltanschauung...the vision of the world. Where you learn how to convey the messages from one idiosyncratic language to the other, you know?'}

\colorbox{traditional-color}{P17} describes that although `there are many people out there who say they're translators', there is potential room for error when the person doesn't have the required experience. This emphasis on translators not only being qualified, but also practicing with care and attention to detail that one gains from experience. A translator's reputation is inherently tied to their work. Therefore, there is a higher risk that a translator with a history of carelessness or lack of contextual awareness `will produce something that's inferior' (\colorbox{traditional-color}{P17}). \colorbox{operationalist-color}{P18} puts a more positive spin on this, saying that a translator's skill-set is valuable and that translation is inherently a difficult task, the precise reason it should be done by a human. They further add that `there are so many things we interact with daily that are translations that we don't realize are translations', emphasizing the benefits everyday speakers of language have from well done human translations.

\subsection{Current Landscape of Translation Technologies}

In this section, we address \textbf{RQ2}. We do this by inquiring about what translation tools are commonly used, where we asked translators directly as part of our interview: \textbf{`What is a translation technology?'}. The responses to this question provide both scale as well as depth into the variety and specificity of different translation technologies currently being used.

Several participants listed several specific translation tools like dictionaries (\freecolor{P1},\tradcolor{P6},\freecolor{P19}), Google Translate (\freecolor{P1}), MemoQ (\opcolor{P3},\opcolor{P4}), and ChatGPT (\freecolor{P1},\opcolor{P7},\opcolor{P14}). This breakdown between physical and non-physical tools is echoed by \colorbox{traditional-color}{P13}, who describes tools as ranging from physical mediums like the ones immediately available to them: `you may see dictionaries behind me [referring to bookshelves in background]' to describing non-physical tools as `the compendium of software or platforms that allow you to get a translation that you can edit or customize for your own use'.

However, even these specific examples cover only a fraction of different ways of assisting in translating \cite{liebling-etal-2022-opportunities}. \colorbox{free-agent-color}{P16} states more broadly:

\begin{quote}
\textit{`Well, technology is as simple as paper and pen.  And as sophisticated as large language models now, or translation memories, or software like Trados. So technology does encompass all of those layers...'}

\hfill ---P16, Free Agent
\end{quote}

Since tools are used in different contexts and domains, the tools afford different outcomes. \colorbox{operationalist-color}{P15} breaks down the different categories of tools:
\begin{itemize}
    \item \textbf{CAT tools: (ex. MemoQ)} These are online datastores of previous translations, with instant lookup. These are useful for structured translation tasks where consistency matters.
    \item \textbf{MT tools: (ex. DeepL)} Translating models that are specifically for translation tasks. Mostly accurate, but limited to specific language pairings and domains.
    \item \textbf{AI tools: (ex. ChatGPT)} Helpful for not just translating, but also more high-level aspects (paraphrasing, rewording, etc.), also better at colloquial language than either CAT or MT tools.
\end{itemize}

This overall approach is echoed by \colorbox{traditional-color}{P19} who also adds `(web)sites that kind of crowdsource as well, like Linguee', acknowledging the various types of databases and search tools that translators use for research.

\subsubsection{Translation tools do not translate, they assist in translation.}
\label{sec:assistive_tools}

Translation technologies are not intended to offer an end all be all solution to translation, instead adhering to `aid in the task of translation' (\colorbox{operationalist-color}{P3}). They do so in various ways, with main emphasis on speed and efficiency.

For many translators like \colorbox{operationalist-color}{P4}, speed and efficiency are some of the main reasons they use tools like LLMs, which they then organize and cross-reference with a CAT tool like MemoQ. CAT tools often have a translation memory component, but they can still differ in specific function. For example, WordFast, used by \colorbox{free-agent-color}{P12} is a paid CAT tool with access to a larger database of translations, while others like SmartCat, used by \colorbox{operationalist-color}{P15}, are free to use in a web-browser.

There is also a difference overall between effectiveness depending on tool type. For CAT tools, an efficient CAT tool was one that prioritized `consistency' and `ease of use' (\colorbox{operationalist-color}{P18}). 
\colorbox{traditional-color}{P13}, who does legal translation, refers to CAT tools as a `clearinghouse' of vocabulary, a collection of readily available terminology for a given use case. For translators that use ML and AI tools in addition to CAT tools like \colorbox{traditional-color}{P2}, they mention how CAT tools are often more restricted in usage and how AI tools like LLMs offer more creative ways to use them.  For MT tools and AI tools, effectiveness is often attributed to speed, as already mentioned, as well as better capturing `tone' and `context' of longer texts (\colorbox{operationalist-color}{P15}). Unlike CAT tools, MT and AI tools are often models trained on much larger data stores, and as such have the ability to generate a larger variety of outputs and process a wider variety of requests. According to \colorbox{traditional-color}{P2}, for AI tools like GPT this comes with both the good and the bad depending on context. For example \colorbox{operationalist-color}{P14} found success in instructing LLMs to adapt the translation in specific ways: 
\begin{quote}
\textit{`I can just simply ask the LLM to: Hey, could you make it shorter? Could you like be more concise?'}

\hfill ---P14, Operationalist
\end{quote}

Whereas others, like \colorbox{free-agent-color}{P16} who tried using LLMs like ChatGPT for religious translations found the outputs to be unsatisfactory: 

\begin{quote}
\textit{`So, ChatGPT for me now is as bland as they come. The output is messy, sometimes inaccurate, and, at best, below average.'}

\hfill ---P16, Free Agent
\end{quote}

Overall, we see the differences in tools and the affordances they lend can vary, but they never replace the translator or the translation task. While some, like LLMs, have come close, there are still user discernments about the output quality.


\subsubsection{LLMs complement, but do not replace existing tools.}

If translation tools have changed how translation works, it therefore follows that the newest development of AI translation tools will similarly alter the translation landscape.

However, what we see is that translators, even when adapting emerging technologies like AI tools into their workflow, are still prioritizing what is important for them: the human aspect. For instance, \colorbox{traditional-color}{P2} uses different LLMs like ChatGPT and Claude in tandem until they `converge' to a ubiquitous answer. While this is not the intended use case of either of these LLMs, \colorbox{traditional-color}{P2} has creatively used them as a tool in a way that more directly benefits their use case. We see how similarly, \colorbox{operationalist-color}{P7} uses LLMs that are present within certain CAT tools as a starting point, but then proceeds to do quality checks and translation on top of it. \colorbox{traditional-color}{P17}, a longtime translator who has observed several `waves' of translation tools, discusses how the introduction of AI tools poses a new, much bigger, obstacle for translators:
\begin{quote}
\textit{`I mean, every generation does this so it's like, yes, there are rules, but humans are still in control.  And so that's (AI tools) where you lose that. You lose that factor when you hand it over to some program  that can't necessarily make judgment calls on certain things.'}

\hfill ---P17, Traditional
\end{quote}

\colorbox{traditional-color}{P10} and \colorbox{operationalist-color}{P4} express similar concerns, and as such only use LLMs on small sentence fragments, and avoid using them on larger texts because they feel they do not have `as much control'.

However, translators have found workarounds for various tools even before the introduction of LLMs. \colorbox{traditional-color}{P8} will default to translating by hand and only use tools when `necessary', usually via an agency when collaborating with other translators. Others, like \colorbox{free-agent-color}{P16}, come up with their own individual alternatives. In doing so, \colorbox{free-agent-color}{P16} simply keeps their own translation memory via an Excel spreadsheet instead of using a CAT tool. They describe this as more straight forward, as their context of translating literary texts in Arabic is very specific and CAT tools do not have the supporting structures to store elements in a organized way. 



\subsubsection{LLMs as a search tool.}

Recently, the development of LLMs has increased in both speed and accuracy for translating across many languages \cite{10.1145/3715336.3735679}. We see how for direct translation tasks, there are often mixed results. For example, \colorbox{operationalist-color}{P11} describes their experience using LLMs to translate into Standard French as fair, but translating into Quebecois or colloquial French as inadequate.

However, when instead used simply as a search tool as opposed to a full translation tool, LLMs have shown to be quite useful for finding and incorporating cultural and linguistic information. For instance, \colorbox{traditional-color}{P8} found success in using Google's AI Studio for finding new vocabulary, and \colorbox{traditional-color}{P10} found that for Sanskrit, Prakrit, and other older Indian languages LLMs are actually better than dictionaries for providing accurate definitions.  

In addition to using an LLM as a search tool, \colorbox{traditional-color}{P2} finds it most helpful when LLMs also provide references with clickable links to specific sources. \colorbox{traditional-color}{P2} also recommends using multiple LLMs in parallel for searching to `maximize search breadth'. These experiences of using LLMs mostly for search tools instead of a full translation pipeline reflect similarly across all translators who have experience using LLMs for translation (\freecolor{P1},\opcolor{P4},\opcolor{P5},\opcolor{P7},\freecolor{P9},\opcolor{P11},\freecolor{P12},\opcolor{P14},\opcolor{15}).



\subsection{Tool \& Translator Misalignments}


In this section, we address \textbf{RQ3} and discuss the attitudes translators have towards certain translation technologies. This includes reasoning for why certain tools are better for certain tasks, as well as why translators prefer certain tools. Finally, we discuss the issues the tools introduce and reinforce, and how these issues impact translators' work approach.

\subsubsection{Translators prioritize their audience understanding, tools prioritize efficiency.}

Throughout our interviews, there is a clear misalignment between goals of translator and translator tools. From our translators, based on the type of decisions they make (Section \ref{sec:decisions}), it is clear that they prioritize the \textit{human} feedback component. However, tools often afford efficiency and time (Section \ref{sec:assistive_tools}). This major misalignment between translator and tool outcome goals results in friction when these differing end-goals transpire in tangential outcomes. This is usually evident in the translation quality from the translators' perspectives.

\colorbox{operationalist-color}{P3}, when doing translation in the website UI design domain, described centering their client needs, and their audience, as one of the main goals for their translation. Similarly, \opcolor{P5},\tradcolor{P6},\tradcolor{P8},\freecolor{P12},\opcolor{P15},\tradcolor{P17}, and \opcolor{P18} all expressed similar sentiments of translating for an `audience', and therefore would make translation decisions to prioritize and ensure the audience's understanding. For instance, \colorbox{operationalist-color}{P14}, who translates UX design interfaces, describes how when translating for a Gen Z audience in particular they would purposefully block, or `ban' words from the available translations within their CAT tool to focus on language that demographic would typically use.

On the other hand, \colorbox{traditional-color}{P13}, who translates legal documents, often times ignores the CAT tool that translates short phrases and simply goes through existing translations or entire documents to see if it can be `recycled', because in law consistency of language is important. In both these scenarios the tools merely are there as an aid, if need be. They are not required to be used in any given instance.


\subsubsection{Translation tools reduce communicative intent.}

One commonality expressed by translators about tools is their inherent ability to dilute language. While CAT tools are by design implementing consistency, AI tools like ChatGPT have a tendency to sound `bland' and `messy' (\colorbox{free-agent-color}{P16}). Conversely, both \colorbox{traditional-color}{P2} and \colorbox{operationalist-color}{P14} agree that ChatGPT often is `too creative' and is often adding words that are not necessary to be there. Even for current topics or slang, \colorbox{traditional-color}{P8} mentions how they often favor Google search instead of ChatGPT since  `something that is trending on TikTok, YouTube, or something, is going to be easier to reach it, via Google than ChatGPT'. 

Outside of AI tools and LLMs, these sentiments of translation technology outputs being mundane or bland are seen in CAT tools as well. For instance, \colorbox{operationalist-color}{P14}, who works in UX translation, cannot use certain words that are `banned words' in CAT tools used in their agency, because those words are not part of the agency's brand voice. Similar restrictions exist in other CAT tools used by agencies \colorbox{operationalist-color}{P15} where the tools must be specialized in order to match the tone of specific phrases and sayings used in product marketing.

Lastly, \colorbox{operationalist-color}{P18} and \colorbox{traditional-color}{P10} both touch upon the chauvinism, specifically within an Anglo-centric context, that often protrudes into translation tools, and as a result, into the translations themselves.


\colorbox{operationalist-color}{P18} discusses how this reflects the translation profession as a whole, where you are more likely to interact with translated content in English mediums:

\begin{quote}
\textit{`I think that people… especially Anglophone people, are like: Oh, but that sullies the language somehow to translate it, that makes it worse... And I think it just makes it different.'}

\hfill ---P18, Operationalist
\end{quote}

These are similarly reflected in the outputs of tools, like for instance when \colorbox{free-agent-color}{P16} describes their experience using LLMs to translate Arabic into English as adequate, but translating from English into Arabic as subpar. 

All of these are examples of how translation tools, while with intent to help, can actually compromise intended meaning and present obstacles for translators to have to address inaccuracies in their work.



\subsubsection{Translators do not trust AI tools.}

Overall, LLMs and AI tools like ChatGPT, Google Gemini, and more have become popular methods for translation rivaling traditional online MT tools like Google Translate and DeepL. Several translators, like \colorbox{traditional-color}{P17}, feel that this is `overhyped', and that these technologies have underlying harms that aren't being openly discussed upfront.

The first of which is simply: translators do not know the underlying mechanisms for how LLMs work. \colorbox{free-agent-color}{P1} comments that they do not use LLMs because they are scared of them and are worried they are collecting their data. \colorbox{traditional-color}{P13}, a legal translator, shares similar concerns and states that they will not use LLMs at any point with translation to avoid any legal confidentiality issues. 

Similarly, other translators have already highly productive and established workflows and adding to it or updating it is more of a nuisance than an aid (\tradcolor{P8},\tradcolor{P6},\freecolor{P16},\tradcolor{P19}). Others are barred from using anything besides the designated CAT tools provided by their agencies (\opcolor{P4},\opcolor{P14}), or are limited by access to internet and digital media (\tradcolor{P8}).

\colorbox{operationalist-color}{P15} describes that the main problem of LLMs, over tools like CAT, is that they have an `increased risk' to be incorrect. \colorbox{operationalist-color}{P3} and \colorbox{traditional-color}{P13} both similarly express encountering occurrences where LLMs have hallucinated or translated terms entirely incorrectly.

Still, others cite outside factors. \colorbox{operationalist-color}{P18} mentions that they don't feel comfortable using LLMs for ethical reasons, citing the harms that LLMs are not environmentally friendly. \colorbox{traditional-color}{P17} even goes further to say that LLM outputs being used for translation can cause spread of misinformation, which if used by the wrong hands can have disasterous consequences if used in domains like law:

\begin{quote}
\textit{`If AI makes a mistake, or DeepL makes a mistake, or Google makes a mistake, you can't hold it accountable.`}

\hfill ---P17, Traditional
\end{quote}

If AI tools were to be utilized more broadly in translation tasks, this type of work would thus require more caution. Instead of translators only having to account for problems like preparation, mentioned in Section \ref{sec:translation-requires-preparation}, they would also need to be constantly making sure their translation is not doing something unexpected. \colorbox{traditional-color}{P13} puts this concern as more inhibitory to their work:

\begin{quote}
\textit{`Ultimately, you're responsible for whatever you render [using AI tools] in that other language, so…  you could use it to draft something, but it's like…  boiling something on the stove. You can't turn your back to it.'}

\hfill ---P13, Traditional
\end{quote}

This distrust of AI tools stems from valid translator concerns about lack of transparency and systemic push of these technologies onto them, without acknowledging or addressing translator needs. As these tools grow increasingly widespread, translators will likely grow more hesitant in using these tools.





\subsubsection{Translators work within complex labor arrangements.}
\label{sec:complex_labor}


Across our participants, the impacts of AI have been felt very differently, in part because those impacts are felt very differently based on the terms and power dynamics in their employment. Even in our sample, we heard directly from translators in a variety of arrangements, both freelance and in house; we also heard secondhand about other arrangements: the disappearing market for one-off translation of academic manuscripts into English in Egypt, for instance (\freecolor{P16}), or the subcontracting of in-house translation work to freelance agencies in the pursuit of scale (\opcolor{P3}). Freelance translators varied widely in the degree of power they were able to exercise in their work, with some (perhaps not coincidentally traditionalists with long careers) like \tradcolor{P6} and \tradcolor{P13} being able to refuse work that did not meet their standards, especially out of concern for their reputation; by contrast, even some in-house translators, like \opcolor{P18}, felt the pressure from their employers to embrace AI: \textit{``Their perspective is, this is the future, this is here, we need to use it so that we can continue to be relevant...couldn't you go so much faster?''}

This push to adopt AI was also a concern for \opcolor{P3}, though in different form. \opcolor{P3} is an in-house translator, but described how their company outsourced much of its work to freelance translators, and how recently, the agencies that they contract to have been submitting identifiably AI-generated text. They then described how LLMs were seen to "solve" the consistency problem created by a freelance, rather than in-house model: \textit{''we don't know who they're using as freelancers. Sometimes we get, like, really good translations, and sometimes they would be really bad, and we couldn't ask them to keep using the same, good freelancer. So, at least with the LLM, it's consistent.''}

Through our interviews, we saw repeatedly how AI translation was often framed as a necessity to keep pace with growing expectations, while translators themselves often saw it as being of an unacceptably low quality that required significant oversight, consistent with the findings in \citet{Woodruff2023HowKW}.







\section{ Discussion \& Conclusion}











Our findings illustrate how the work of translation cannot be separated from its inherently human nature. They highlight how the dual utility of translation technologies as powerful machines that can refine the skills needed to produce consistent translations while simultaneously challenging the control that translators have over their decision-making processes. Participants discuss the importance of familiarity with texts, notably for the translation of literary texts (Section \ref{sec:translation-requires-preparation}). Translation work is a skill deeply rooted in continuous and iterative practices that define the unique stylistic approach of a translator. For many of the participants we interviewed, their stylistic approaches were refined over time and protected through a commitment to the craft of translation. 
Our participants purport, in alignment with the humanistic nature of translation work, that translation is a human-centered practice. 

In this discussion section, we outline three key elements that illustrate how translation technologies transform the nature of translation work. First, we argue that the introduction of translation tools in translation work further highlights just how dependent translation work is on \textit{craft}, the situated and tacit knowledge-based skills of translators. Secondly, we discuss how translation tools, if used incorrectly, not only contribute to the erasure of the craft of translation but also both decontextualizes and reduces the quality of the translations themselves, which is a key aspect of translation as described by our study participants. 
Finally, we discuss the friction that translation technologies present in the division of labor as it relates to the professional field of translation. Ultimately, translation technologies disrupt normative approaches to translation that, for many, are deeply personal, subjective, and until recently, a creative practice that reflects the decision-making of the translator. We conclude our discussion by highlighting the implications of the friction between experts and the technologies that aid their work.    

\subsection{Translation Practice as Humanistic Craft}
\label{sec:human-discussion}

Translation practices are a type of creative skill that reflects the interpretive style of a translator. Our participants described their translation work as experiential, where judgment calls on providing the communicative meaning of translation relies on personal judgment. \colorbox{traditional-color}{P17} describes how translation is not simply a practice of translating words but feelings as well: \textit{`you're translating feelings, not just words'}. Participants' commitment to research on the context and domain of the translation illustrates just how influential and iterative translation work can be. Translation is more than a replication of words from a source language to a target language, rather it is a stylistic practice that conveys the meaning behind those words while considering the linguistic features of the source language. Translators, in turn, consider the problem of translatability, the process of accurately communicating meaning, as complex and highly nuanced \cite{sun2012translatability}. This requires judgment calls on how to describe concepts that cannot be readily defined in another language. For example, \colorbox{operationalist-color}{P15} discussed improvising the description of a `Hot Pocket' as it would be described in Chinese. Improvisation and the will to find connotative meaning beyond literalness illustrate the humanistic nature of translation work, a feature of a translator's expertise that was widely discussed in our study. 

The perspective of translators is also deeply situated and can be described as co-produced from the lived-experiences of the translators. This kind of situated knowledge \cite{haraway2013situated} is a key component of a human-centered translation process. Our participants recognized the work of contextualization as a creative practice that relies on the perspective of the translator. Furthermore, a `good' translation requires creative interventions that have long been recognized by scholars in the field of translation studies \cite{grass2023translation}. The incorporation of translation tools into translation work offers a reminder of how the most important components of translation work require the subjectivity of human reasoning. 

\subsection{Labor Relations Mediate Use of Translation Tools}
\label{sec:infringing-discussion}

Translators in our study find use for generative AI-based tools. However, echoing findings from \citet{rivasginel_translators_2025}, the utility they find in AI-based tools is often not directly in the work of translation itself, but rather for related tasks like understanding context, or serving as a preliminary search tool to guide access to more traditional resources.

We primarily see these types of use, however, in cases where translators are free to exercise some degree of their own agency in when, or indeed whether, to use the tools. We also see cases that reflect broader trends in the industry like the move towards machine translation post-editing (MTPE) as discussed in e.g. \citet{merchant2025ai}, or documented in the European Language Institute Survey (ELIS) of translation professionals \citep{elis_european_2025}. Rather than being deployed by translators in ways that are compatible with their work, AI translation tools are often deployed in ways that reduce the scope of a translator's involvement. The ELIS survey, for instance, documents that MTPE work for translators often comes from a version of the text that has been pre-translated by a client using an AI tool \citep[][p.35]{elis_european_2025}. In our study, participants like \colorbox{operationalist-color}{P18} discuss feeling the pressure to use AI to accelerate their work in ways that were not compatible with their quality standards for translation. In this view, translation is transformed from an integrated practice of transforming meaning to digital piece work, in which their compensation is tied to discrete, decontextualized units of work, and much more reliant on the quantity, rather than the quality of the translated work. 

This transformation, from translators being responsible for integral units of work to decontextualized snippets, reflects a minimization of the expertise that translators build and value. Instead, the machine-generated translation is viewed as mostly complete, and the translator's agency is reduced to ``fixing" the work. In the piece work metaphor, sewing an individual seam rather than constructing a garment. This claim in many ways parallels the experience of data laborers like crowd workers for Amazon Mechanical Turk, or those who annotate data for AI. As \citet{irani_cultural_2015} argues, framing these workers' labor as small tasks, like ``fixing" machine output, serves to legitimize narratives of technological capability and profit-driven progress, in this case, stabilizing the view of machine translation as the primary source of truth. Thus, a translation might be more work than producing it, and results in what is widely perceived as a lower quality translation \citep[][p.34]{elis_european_2025}. 

\citet{roberts_your_2021} argues that the figure of the ``automated" process, despite its reflection of human judgment, can often serve rhetorical purposes, as in the case of moderation algorithms, where automatic moderation confers a sense of objectivity, and thereby legitimacy. Similarly here, a machine translation, which need only be fixed, provides the justification to reconfigure the labor relations under which translation work can be done, a process which \citet{moorkens_tiny_2020} likens to a digital version of Taylorism, in which tasks are broken down into surveilled, standard units that are seen as being completable by commoditized, non-expert labor. Translation ``expertise", is in this view centralized as machine judgment, with a corresponding casualization and de-skilling of the experts \citep{sambasivan_everyone_2021}. This follows with Section \ref{sec:complex_labor}, where translators expressed concerns over their work being outsourced to AI tools, as well as non-experts who use AI tools with less regulation and without the proper expertise.

\subsection{Resisting The Erasure of Craft}
\label{sec:craft-erasure}

Overwhelmingly, translators consider their work as a type of craft, a socially embedded project that relies on cultural understanding and cultivated expertise to balance competing values in the translations they produce. By contrast, the tools that are available to them are often designed with a much shallower view of translation 
that often precludes the rich, contextual reasoning necessary for what they consider good translation. In addition, these tools facilitate a reorganization of labor that minimizes translators' carefully built expertise and alienates them from their labor \cite{miceli2025methodological}. 

Despite these costs, however, our participants remain divided on the use of AI tools. While some of our participants recognize the potential labor impacts of AI and reject it, similar to the visual artists surveyed in the work of \citet{jiang2026professional}, AI tool use remains common in our participant pool. This divide emphasizes that while AI tools have potential for assistive use in translation, that potential is often realized despite the design of these tools. The translators in our sample rarely use LLM-based tools for the task of translation itself, instead relying on them for access to contextual information, fuzzy retrieval, or to fill out a distribution of possible translations to better consider options. These patterns of use outline design directions for AI tools that preserve and augment the expertise of translators as well as the quality of translations. More purpose built tools might, for instance, suggest a range of possible translations, or provide exploratory interfaces with links back to historical corpora rather than suggest candidate translations to be ``fixed". These design directions acknowledge and center the uncertainty of translation and the trained discretion of translators. 

\subsection{Conclusion}

In this study, we explore the role of translation technologies in how translators engage in translation work. As translators recall their experiences working and contending with these technologies, they describe their work as craft. Efforts in protecting the craft of translation are vulnerable in multiple directions. As discussed, threats of deskilling and the reorganization of labor complicate translators' trust in translation technologies. However, it remains painfully apparent that market forces and top-down decision-making position translators to face an uncertain future with increasingly precarious work conditions. Tools like LLMs complicate this dynamic by normalizing a cultural view of translation that devalues the expertise and judgment of translators, which in turn facilitates that economic reconfiguration. 

Echoing \citet{diaz_what_2024}, we therefore call for the communities developing translation tools to more deeply engage with the expertise of translators by acknowledging the limitations of tools to consider the craft of translation, and to build tools that center translators' epistemic authority. We assert that lessons from translators' perspectives provide useful insight into other types of domain expertise that are increasingly facing disruptions due to AI. This study illustrates the importance of seeing expert-based work as craft work.

\subsection{Limitations}

This study by no means represents the viewpoints of all professional translators. Since we recruited with convenience and snowball sampling, our sample of participants was inherently scoped to those close to or adjacent to academia. Similarly, our pool of languages and domains covered was limited to those seen in Tables \ref{tab:age and domains} and \ref{tab:languages}. Any other languages or domains were not represented in these findings, especially those from low-resourced linguistic backgrounds. The 19 interviewees instead serve as a glimpse into what perspectives and conversations exist around translation technologies.

In this work, translators self-identified as professional translators. This is because there does not exist just one standardized `professional translator' test or evaluation \cite{galan2015competence, li2020self}. We determined if they were eligible simply by their own discretion, as well as their responses to the form questions `Yes' to the question: \textbf{`Do you have translation certification (a degree, certificate, or other proof of translation training)?'}.

\section*{Acknowledgements}

We would like to thank Shaily Bhatt, Anjali Kantharuban, Leena Mathur, Joel Mire, Malia Morgan, Jimin Mun, Ollie Radović, Jocelyn Shen, and Kynnedy Smith for their help and feedback on prior drafts of this work.

This study was funded by the Portuguese Recovery and Resilience Plan through project C645008882-00000055 (i.e., the Center For Responsible AI) and by FCT/MECI through national funds and co-funded EU initiatives under UID/50008 for Instituto de Telecomunicações.

\bibliography{aaai2026}

\end{document}